# An analysis of derivative absorption spectroscopy of multiferroic bismuth ferrite materials

**Ramachandran Balakrishnan[1,*]**

[1]Physics Research Centre, Department of Physics, Sethu Institute of Technology, Pulloor, Kariapatti-626115, Virudhunagar, Tamil Nadu, India.

[*]Email: ramskovil@gmail.com

## ABSTRACT

The probability density function (PDF) and cumulative density function (CDF) of bulk $BiFeO_3$, nanostructured $BiFeO_3$, and thick film of $BiFeO_{2.85}$ were studied in detail based on their experimental absorption data. The goal of this study was to investigate the electronic transitions and the Urbach tail band in three samples of the magnetoelectric multiferroic $BiFeO_3$. The PDF and CDF functions were derived from the absorption data of these substances in the UV-visible-NIR region of light spectrum. For the $BiFeO_3$-based materials, charge transfer (*p-d* and/or *p-p*) transitions in the energy range of 2.5 eV to 5.5 eV were identified through PDF and CDF analysis. Furthermore, spin-orbit and electron-lattice interactions along with a low-symmetry crystal field cause two doubly degenerate *d-d* transitions of $Fe^{3+}$ ions in the $FeO_6$ octahedra of the $BiFeO_3$ samples to occur between 1.5 eV and 2.5 eV. Additionally, the $BiFeO_3$-based materials exhibited strong direct and weak indirect transitions near the band edge, which point to a complex band structure of $BiFeO_3$. Finally, the energy of the defect-induced Urbach tail band was directly calculated by examining the CDF function of the materials, yielding estimated values of approximately 0.40 eV, 0.31 eV, and 0.33 eV for the bulk, nanostructured, and film $BiFeO_3$ samples, respectively.

1. Introduction

Optical spectroscopy is a non-destructive technique commonly used to investigate the optoelectronic characteristics of materials by measuring the absorption, reflection, and emission of light as it interacts with matter in the optical region of electromagnetic spectrum, which ranges from 200 nm to 3000 nm [1-4]. Specifically, UV-visible optical spectroscopy offers insights into the electronic structures of materials [4] and reveals the optical transitions within materials from filled to unfilled states. The term absorption spectroscopy refers to optical spectroscopy that focuses on absorption measurements. Absorption occurs when materials interact with light via various processes [2]. Transition metal ions like Ti, V, Cr, Mn, Fe, Co, Ni, and Zn in their different oxidation states induce light absorption in their host materials in various mineral forms. Generally, the optical absorption spectra of mineral materials that include transition metal ions are affected by six processes [2,5]. These processes involve electronic transitions between the valence band maximum (VBM) and conduction band minimum (CBM), electronic transitions that include the displacement of charge density from one ion to another (known as charge transfer transition), and electronic transitions that engage electrons in $d$ orbitals of first-row transition metal ions such as $Cr^{3+}$, $Mn^{2+}$, $Fe^{2+}$, $Fe^{3+}$. These three processes are primarily observed in transition metal-containing oxides. Since valence electrons in semiconducting and insulating materials are bound electrons, the transitions related to them are termed interband transitions [2,6].

Tanabe and Sugano [7,8] provided a comprehensive explanation using diagrams named after them regarding how the participation of $d$-orbitals leads to $d$-$d$ transitions that impart color to mineral materials consisting of third-row transition metal ions like Ti, V, Cr, Mn, Fe, Co, and Cu. In particular, the strong charge transfer transitions are primary contributors to the absorption spectra of materials with $Fe^{3+}$ ions and their low-energy branches appear in the visible and near-infrared (NIR) regions [9]. Conversely, the $d$-$d$ transitions of $Fe^{2+}$ predominantly take place in the infrared region [10,11]. Recent studies on multiferroic $BiFeO_3$ have disclosed that spin-orbit coupling relaxes spin-forbidden two-doublet $Fe^{3+}$ $d$-$d$ excitations in octahedral symmetry within the energy spectrum of 1.0 eV to 2.0 eV [12-14]. The splitting of the triply degenerate $^4T_{1g}$ and $^4T_{2g}$ electronic energy levels due to the symmetry reduction from $O_h$ to $C_{3v}$ in BiFeO$_3$ accounts for the two doubly degenerate electronic transitions with *A* and *E* characteristics that have been experimentally detected [12,14].

Over the past few decades, bismuth ferrite (BiFeO$_3$) has been the focus of much research for various applications, particularly in the energy sector, including photovoltaic and photocatalytic devices, among others [15-20]. Because of its room-temperature multiferroic properties, which include ferroelectricity, antiferromagnetism, and ferroelasticity, BiFeO$_3$ stands out as a material [20-23]. This material has attracted a large interest among scientists and engineers due to its lower band gap energy ($E_g$) of approximately 2.5–2.7 eV reported for BiFeO$_3$ [12-14,24-28], which is less than that of conventional ferroelectric materials such as BaTiO$_3$ and Pb(Zr, Ti)O$_3$ ($E_g$ >3 eV). Consequently, a significant amount of research has been done on BiFeO$_3$ to examine its photoelectric properties, such as photovoltaics, photocatalysis, and photoferroelectrics [15-18,29]. However, the energy band gap of this material is often miscalculated due to the dominance of absorption at lower energies than expected [30,31]. The primary reason for this is the presence of Urbach tails or bands caused by structural defects, thermal excitations, and other factors [30-33]. To examine the electronic transitions and defect-associated Urbach energy in bulk, film, and nanostructured forms of BiFeO$_3$, this study explores a novel method called derivative absorption spectroscopy analysis on the absorption spectra of the BiFeO$_3$-based materials. The purpose of this investigation was to compare the energy of various observed electronic transitions and the electronic structure of the model magnetoelectric multiferroic material BiFeO$_3$ in bulk, nanostructured, and thick film forms.

## 2. Method and Materials

The absorption spectra, *α(E)* of bulk and nanostructured BiFeO$_3$ samples within the energy range of 1.5 eV to 5.5 eV were derived from the relative reflectance data by employing the Kubelka-Munk theory. The relative reflectance ($R = \frac{R_{sample}}{R_{standard}}$) measurements for both of these samples were taken using the Ocean Optics Spectrometer (USB2000, USA), as detailed in a recent study by Ramachandran *et al.* [34]. The absorption spectrum, *α(E)* of oxygen-deficient BiFeO$_{2.85}$ thick film (with a thickness of 2 $\mu$m) was computed using its measured reflectance (*R*) and transmittance (*T*) spectra through the relationship of absorbance (*A*) with the parameters *R*, *T*, and *α* [35]:

$$A = \alpha d = -\ln\frac{(1-R)^2}{T}. \qquad (1)$$

Where *d* is the thickness of the film. Further information regarding the measurements of *R* and *T* on the studied thick film can be found in Ref. 35. The preparation and characterization of the inspected $BiFeO_3$-based bulk, film, and nanostructured samples were studied and documented in previous reports by Ramachandran *et al.* [14,22,27,34-38].

The absorption spectra, *α(E)* of bulk, film, and nanostructured multiferroic $BiFeO_3$ materials were examined using a novel analytical technique introduced by Canul *et al.* [31]. This technique employs derivative absorption spectroscopic analysis along with deconvolution methods. Comparisons between findings from this analysis and conventional optical spectroscopic analysis, such as the Tauc plot, Urbach analysis, and the first derivative of reflectance ($\frac{dR}{d\lambda}$) [34,35], were made. The utilized analytical technique aims to remove the signal arising from artifacts in the *α(E)* spectra of the samples. Furthermore, this approach will prove to be very useful for analyzing materials with intricate band structures [31], as the Urbach rule has not been formulated for the nonlinear band edges exhibiting a complex fine structure. However, it is used as a general rule in the evaluation of optical properties and electronic structures of a variety of semiconducting and insulating materials [31,34,39-46]. The Urbach analysis employs a graph of the logarithmic absorption coefficient as a function of energy, $\ln \alpha (E)$ to derive the linear nature of the Urbach rule [31,47-49],

$$\ln \alpha (E) = \ln a_0(E) + \frac{E-E_o}{E_U}. \qquad (2)$$

Where the Urbach energy ($E_U$) is inversely proportional to the slope, $\frac{d \ln \alpha(E)}{dE} = \frac{1}{E_U}$, and $a_0$ is a scaling parameter and $E_o$ is the Urbach focus. The Urbach energy is considered a measure of the broadening occurring at the band-edge of semiconducting materials and the softening or decrease in their slope, $\frac{d \ln \alpha(E)}{dE}$ at the band-edge [47,48].

Pankove [30] demonstrated that the absorption coefficient is directly related to the cumulative or integral probability of transitions between energy states. Recently, Canul *et al.* [31] indicated that the experimental signal component ($\frac{d \ln \alpha(E)}{dE}$) corresponds to a probability density. In other words, the derivative of a cumulative probability function (CDF) is equivalent to the probability density function (PDF) in statistics. This reveals the normally distributed variability in the local potential caused by the random distribution of induced defects in the materials according to semiconductor theory [47,48]. The experimental signal component $\frac{d \ln \alpha(E)}{dE}$ serves as a good Gaussian approximation to a simpler logistic distribution of probability density function [31,32],

$$\frac{d \ln \alpha(E)}{dE} = \frac{a_c e^{\frac{-(E-E_o)}{\gamma}}}{\gamma \left(1+e^{\frac{-(E-E_o)}{\gamma}}\right)^2}, \qquad (3)$$

Here, $\alpha_c$ represents the total quantity of absorbed photons or the relative photon capacity and corresponds to the integral area of signal peaks. The factor $\gamma$ is related to the dispersion in energy close to the band-edge due to fluctuations in the local potential associated with induced defects [31,47,48]. Interestingly, the Urbach energy can be evaluated from the value of the inverse slope of the $\frac{d \ln \alpha(E)}{dE}$ data versus $E$ (i.e., CDF function) at the inflection point of $E_o$. These differential analyses are very useful methods for quantitatively analyzing the characteristic behavior (i.e., slope feature) of different elements of the complex band-edge function [49]. Therefore, in this work, the absorption spectra of the bulk, thick film, and nanostructured $BiFeO_3$ samples were probed using a derivative logarithmic function of the absorption coefficient with energy ($E$) as described in equation (3). The presented analyses were carried out using Origin scientific graphing software, while the line shape analysis of the absorption spectra presented in this study was conducted by employing an interference pattern deconvolution technique using the PeakFit software package [31,32].

## 3. Results and discussion

### *3.1. Analysis of charge transfer transitions and band-edge in bulk BiFeO$_3$*

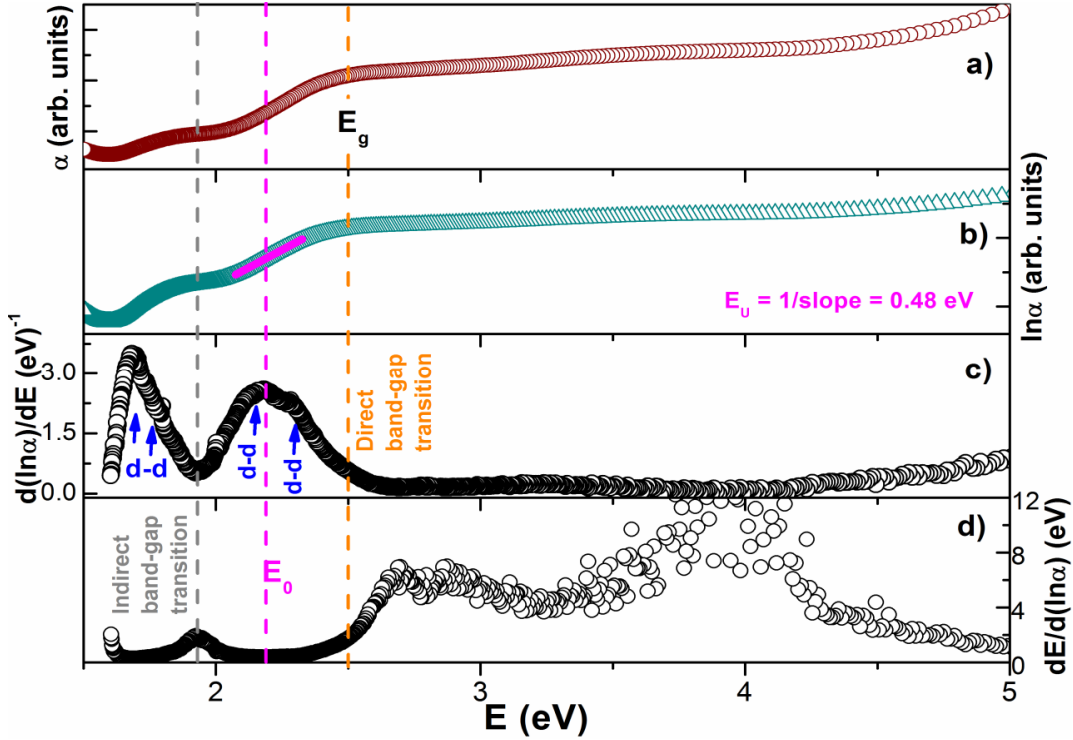

**Fig. 1** The energy-dependent curves of a) absorption coefficient ($\alpha$), b) the logarithmic absorption coefficient ($ln\alpha$), c) the energy derivative of the logarithmic absorption coefficient, $\left(\frac{d \ln \alpha(E)}{dE}\right)$, and d) the inverse term $\left(\frac{dE}{d \ln \alpha(E)}\right)$ of bulk BiFeO$_3$.

The graphs of the energy-dependent absorption coefficient ($\alpha$) and the logarithmic absorption coefficient ($ln\alpha$) for bulk BiFeO$_3$ in the range of 1.5 eV to 5.0 eV are presented in Figs. 1a and 1b, respectively. Two distinct features related to the optical band gap edge and the *d-d* transitions of Fe$^{3+}$ ions within the octahedral arrangement of BiFeO$_3$ were observed below 2.6 eV, indicated roughly by dashed vertical lines at 2.5 eV and 1.9 eV. The $\alpha$ value shows a significant decrease below 2.5 eV due to the anticipated optical band-edge along with a cusp-like characteristic evident below 2 eV attributed to the *d-d* transitions of Fe$^{3+}$ ions in BiFeO$_3$. The band gap energy of bulk BiFeO$_3$ was determined to be approximately 2.5 eV using methods such as Tauc plot analysis and the first derivative of reflectance [34]. The best linear fit (represented by a solid line in Fig. 1b) to the *ln$\alpha$(E)* plot near the optical band-edge yielded the defect-induced Urbach energy ($E_U$) of about 0.48 eV for bulk $BiFeO_3$ (Table 1). These findings were further investigated using derivative absorption spectroscopy as outlined in equation (3) in the following.

The energy derivative of the logarithmic absorption coefficient, $\frac{d \ln \alpha(E)}{dE}$, which is related to the PDF function for bulk $BiFeO_3$ was calculated using equation (3) and is shown in Fig. 1c. The reciprocal term $\left(\frac{dE}{d \ln \alpha(E)}\right)$ that relates to the CDF function of bulk $BiFeO_3$ was also evaluated and is presented in Fig. 1d. Distinct features associated with charge transfer and *d-d* transitions, as well as the features of direct and indirect band gap transitions (indicated by dashed lines), were observed in Figs. 1c and 1d. To explore these features thoroughly, the deconvolution technique using the Gaussian approximation [31,32] was applied to probe all individual transitions in the $\frac{d \ln \alpha(E)}{dE}$ data with a goodness of fit ($R^2$ = 0.995), as illustrated in Fig. 2. The charge transfer transition related to the dipole-allowed *p-d* transition in $BiFeO_3$ was located at 3.16 eV which is visible in the inset of Fig. 2. The transition associated with the strong direct band gap transition was found at $E_g$ = 2.46 eV (Table 1), consistent with prior studies of Ramachandran *et al*. [14,34]. Moreover, four transitions at 1.67, 1.76, 2.15, and 2.32 eV were detected which are attributed to the doubly-degenerate *d-d* transitions of $Fe^{3+}$ ions in the $FeO_6$ octahedra within $BiFeO_3$ [14,26]. The energies of all the detected transitions in bulk $BiFeO_3$ closely match the values obtained through the methods of the first derivative of reflectance, the Kubelka-Munk function, and the Tauc plot analysis (see Table 1) [34].

Table 1 The determined direct band gap and Urbach energies of the investigated $BiFeO_3$-based materials using their PDF and CDF functions.

| Sample | $E_g$ (eV) using the first derivative of reflectance [34] | $E_g$ (eV) using the PDF function [This work] | $E_U$ (eV) using the Urbach relation [34] | $E_U$ (eV) using the CDF function [This work] |
|---|---|---|---|---|
| Bulk $BiFeO_3$ | 2.48 | 2.46 | 0.48 | 0.40 |
| Nanostructured $BiFeO_3$ | 2.61 | 2.47 | 0.35 | 0.31 |
| $BiFeO_{2.85}$ thick film | 2.27 | 2.32 | 0.38 | 0.33 |

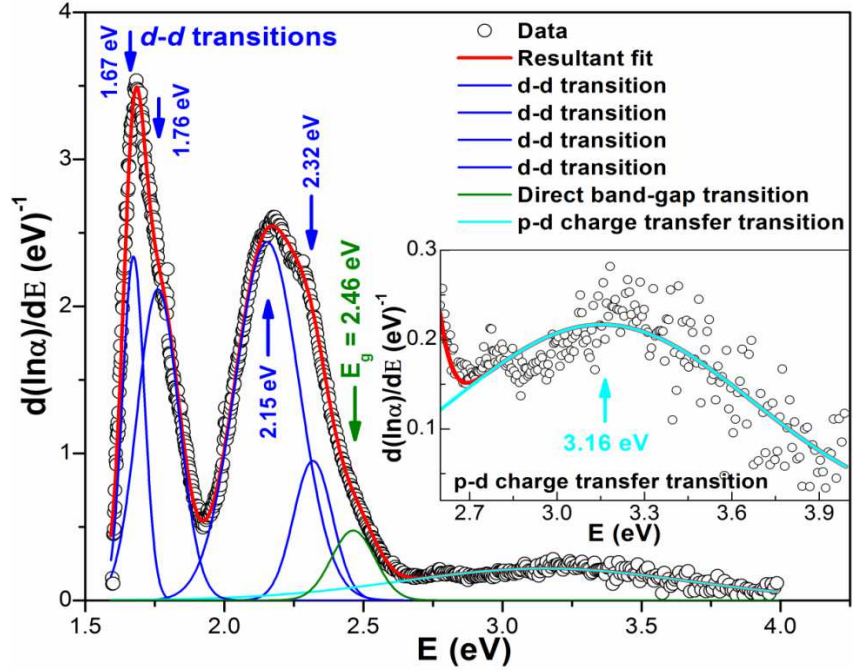

**Fig. 2** The deconvoluted energy derivative of the logarithmic absorption coefficient, $\left(\frac{d \ln \alpha(E)}{dE}\right)$ data of bulk $BiFeO_3$ with the *p-d* charge transfer transition (cyan curve), direct band gap transition (olive curve), four *d-d* transitions (blue curves), and the resultant fit (red curve).

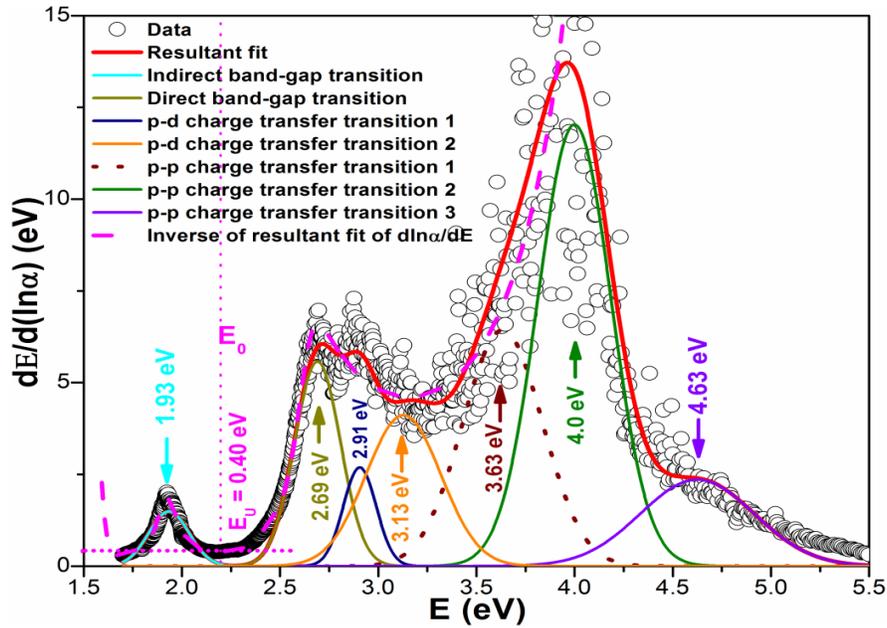

**Fig. 3** The deconvoluted the inverse term $\left(\frac{dE}{d \ln \alpha(E)}\right)$ related to the CDF versus energy (*E*) for bulk $BiFeO_3$. The fitted individual peaks related to the indirect and direct band gap transitions, two *p-d* charge transfer transitions, and three *p-p* charge transfer transitions along with the resultant fit were plotted with the experimental data (circular symbol).

The deconvoluted data for the reciprocal of the energy derivative of the logarithmic absorption coefficient, known as the CDF function $\left(\frac{dE}{d\ln\alpha(E)}\right)$, of bulk BiFeO$_3$ is shown in Fig. 3. Using the deconvolution analysis, the CDF data was accurately fitted with seven components of the complex band structure of BiFeO$_3$ [50]. These components include one in-band gap transition associated with the indirect band gap transition, two *p-d* charge transfer transitions, three *p-p* charge transfer transitions, and one transition related to the strong direct band gap transition. At 2.91 eV and 3.13 eV, two *p-d* charge transfer transitions were observed, which are in line with the values (2.9 eV and 3.15 eV) that Pisarev *et al.* [26] previously reported for the polished (001)-type surface of BiFeO$_3$. The observed *p-d* charge transfer transition at 3.13 eV is consistent with both the PDF function-derived value (3.16 eV) (see Figs. 1c and 2) and the value (3.1 eV) obtained from the $\left(\frac{dR}{d\lambda}\right)$ spectrum of the same bulk BiFeO$_3$ [34].

The CDF data, $\left(\frac{dE}{d\ln\alpha(E)}\right)$ (Fig. 3) revealed three peak features at 3.63 eV, 4.0 eV, and 4.63 eV, which are ascribed to the *p-p* charge transfer transitions in BiFeO$_3$. This finding is consistent with the results obtained from the $\frac{dR(\lambda)}{d\lambda}$ spectrum by Ramachandran *et al.* [34]. The energies (4.0 eV and 4.63 eV) of two of these *p-p* transitions in bulk BiFeO$_3$ are consistent with the values (3.95 eV, and 4.54 eV) of the (001)-oriented BiFeO$_3$ film [26]. The direct band gap transition in bulk BiFeO$_3$ also causes another peak feature at 2.69 eV with the optical band-edge at around 2.4 eV. Additionally, the CDF spectrum clearly shows a peak feature at 1.93 eV, likely due to a weak indirect transition in BiFeO$_3$. This value is similar to the estimated indirect band gap energy (1.86 eV) of the same sample using the Tauc plot analysis [34]. Interestingly, the drawn slope along the *x*-axis (the dotted magenta line in Fig. 3) at the Urbach focus in the CDF function, $\left(\frac{dE}{d\ln\alpha(E)}\right)$ directly yields the defect-induced Urbach energy of about $E_U = 0.4$ eV (Table 1). This value is comparable to the value of $E_U = 0.48$ eV obtained using the plot of *lnα(E)* vs *E* (see Fig. 1b). The inverse of the resultant fit of the PDF function, $\frac{d\ln\alpha(E)}{dE}$ is shown by the dashed magenta line in Fig. 3, which also has a slope with the value of $E_U = 0.4$ eV at the inflection point of the Urbach focus [31].

## 3.2. Analysis of charge transfer transitions and band-edge in nanostructured $BiFeO_3$ sample

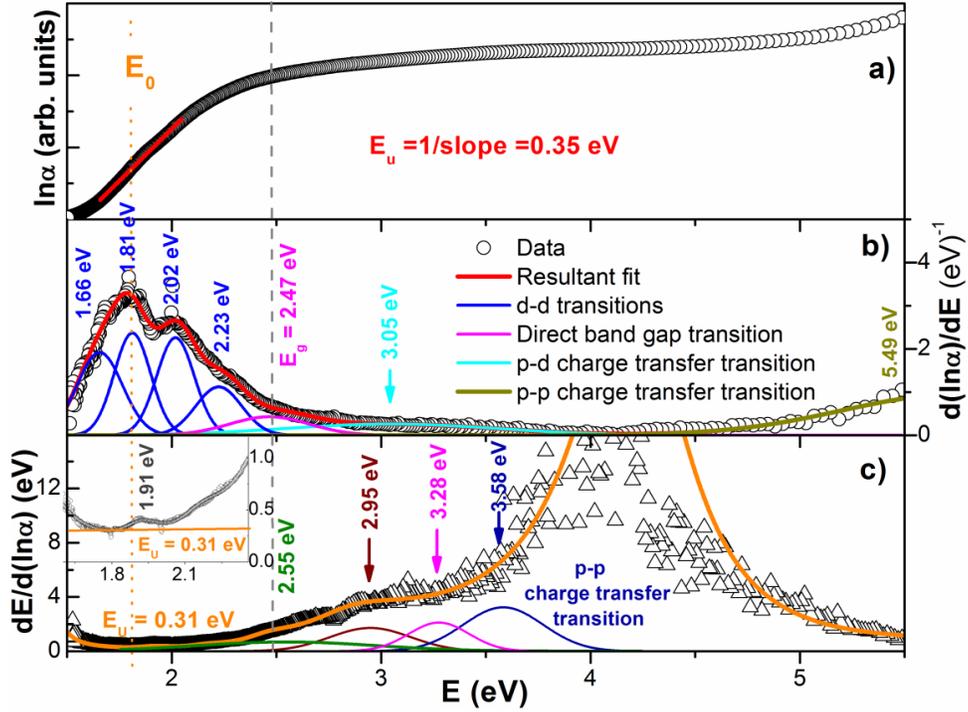

**Fig. 4** a) The energy-dependent logarithmic absorption coefficient ($ln\alpha$) of the nanostructured $BiFeO_3$ sample, b) its logarithmic absorption coefficient, $\left(\frac{d\ln\alpha(E)}{dE}\right)$ and c) the corresponding inverse term, $\left(\frac{dE}{d\ln\alpha(E)}\right)$.

The logarithmic absorption coefficient ($ln\alpha$) of the nanostructured $BiFeO_3$ sample is shown in Fig. 4a within the energy range of 1.5 eV to 5.5 eV. The absorption coefficient ($\alpha$) data for the sample, as provided by Ramachandran *et al.* in Ref. 33, was used to generate this data. The $E_U$ value was found to be about 0.35 eV (Table 1) by fitting a linear regression to the $ln\alpha(E)$ plot, which is associated with the induced structural defects in the nanostructured $BiFeO_3$ [34]. Figures 4b and 4c illustrate the resulting PDF and CDF functions for the nanostructured $BiFeO_3$ sample using equation (3). The characteristics related to charge transfer transitions and *d-d* transitions of the $Fe^{3+}$ ions in the $FeO_6$ octahedra of the nanostructured $BiFeO_3$ sample were observed in the deduced PDF and CDF data (Figs. 4b and 4c). For the inspected $BiFeO_3$ nanostructured sample, two charge transfer transitions, in addition to the features of the optical band-edge features and four *d-d* transitions, were distinctly identified by employing the deconvolution technique with a Gaussian approximation to the PDF data [31,32].

Moreover, two noticeable peak features at 3.05 eV and 5.49 eV (Fig. 4b) were observed, which correspond to the dipole-allowed *p-d* transition and *p-p* charge transfer transition of rhombohedral BiFeO$_3$, respectively. The observed peak feature at 2.47 eV (Table 1) is essentially due to the direct band-gap transition of the nanostructured BiFeO$_3$ sample. This value is consistent with the values (2.40–2.61 eV) obtained for this sample using the Tauc plot analysis and the $\frac{dR}{d\lambda}$ data [34]. Four doubly-degenerate *d-d* transitions detected at 1.66, 1.81, 2.02, and 2.23 eV for the nanostructured sample were consistent with the results of bulk BiFeO$_3$ (see section 3.1) and also with the literature [14,26,27,34]. Similarly, the three peak features at 2.95 eV, 3.28 eV, and 3.58 eV in the CDF function of the nanostructured BiFeO$_3$ (Fig. 4c) correspond to one *p-d* and two *p-p* charge transfer transitions of Fe$^{3+}$ ions in FeO$_6$ octahedra in BiFeO$_3$ [14,26]. The energies of these transitions are well matched with the results of bulk BiFeO$_3$ (see Fig. 3). Meanwhile, the direct band gap-related peak feature is seen at $E_g$ = 2.55 eV in the CDF function, which is consistent with the obtained $E_g$ value of 2.47 eV using the PDF function of the nanostructured BiFeO$_3$ sample. Additionally, a weak peak near 1.90 eV is observed in the CDF data (see the inset of Fig. 4d), which is ascribed to a weak indirect transition in the nanostructured BiFeO$_3$. Importantly, the Urbach energy related to the induced defects is estimated to be approximately 0.31 eV by drawing the slope at the inflection point of the Urbach focus (see the inset of Fig. 3c), which is consistent with the value (0.35 eV) obtained using the *lnα(E)* plot for this nanostructured sample (see Table 1).

### 3.3. Analysis of electronic transitions and band-edge in oxygen-deficient BiFeO$_{2.85}$ thick film

Figures 5a and 5b display the energy-dependent absorption coefficient (*α*) and the natural logarithm of the absorption coefficient (*lnα*) of BiFeO$_{2.85}$ thick film within the range of 1.5 eV to 4.0 eV, respectively. From the *lnα(E)* graph, the $E_U$ value for the thick film is found to be about 0.38 eV (see Table 1). A comprehensive analysis and discussion of these graphs were recently published by Ramachandran *et al.* [35]. The PDF function, $\left(\frac{d \ln \alpha(E)}{dE}\right)$ for BiFeO$_{2.85}$ thick film was derived by differentiating *lnα* with respect to *E*, as illustrated in Fig. 5c. Subsequently, the deconvolution method with a Gaussian approximation was applied to estimate the energies of the observed peak features in the PDF function [31,32]. One *p-p* charge transfer transition was identified at 3.20 eV (see the inset of Fig. 5c), which is very close to the values (3.05–3.28 eV) of both bulk and nanostructured BiFeO$_3$ materials (see Figs. 2–4). Using the Tauc plot analysis in conjunction with a baseline method (see S1 in the supplementary material), an optical absorption edge was noted at $E_g$ = 2.32 eV (see Table 1), which is consistent with its direct band-gap value of $E_g$ = 2.39 eV. The oxygen-deficient BiFeO$_{2.85}$ thick film has a red-shifted band gap energy when compared to bulk BiFeO$_3$ ($E_g$ = 2.46 eV). This finding is essentially due to the oxygen vacancies causing electronic shifts in the constituent atoms of the BiFeO$_{2.85}$ film [35]. Moreover, the near-edge absorptivity ratio ($NEAR = \frac{\alpha(E_g)}{\alpha(1.02E_g)}$ [52]) for the BiFeO$_{2.85}$ film is estimated to be approximately 0.8, which is slightly lower than the value (0.96) for the examined bulk BiFeO$_3$ sample [34]. Thus, the BiFeO$_{2.85}$ film has a lower $E_U$ value of 0.38 eV than the bulk BiFeO$_3$ ($E_U$ = 0.48 eV), which was calculated using the Urbach empirical formula given in equation (2).

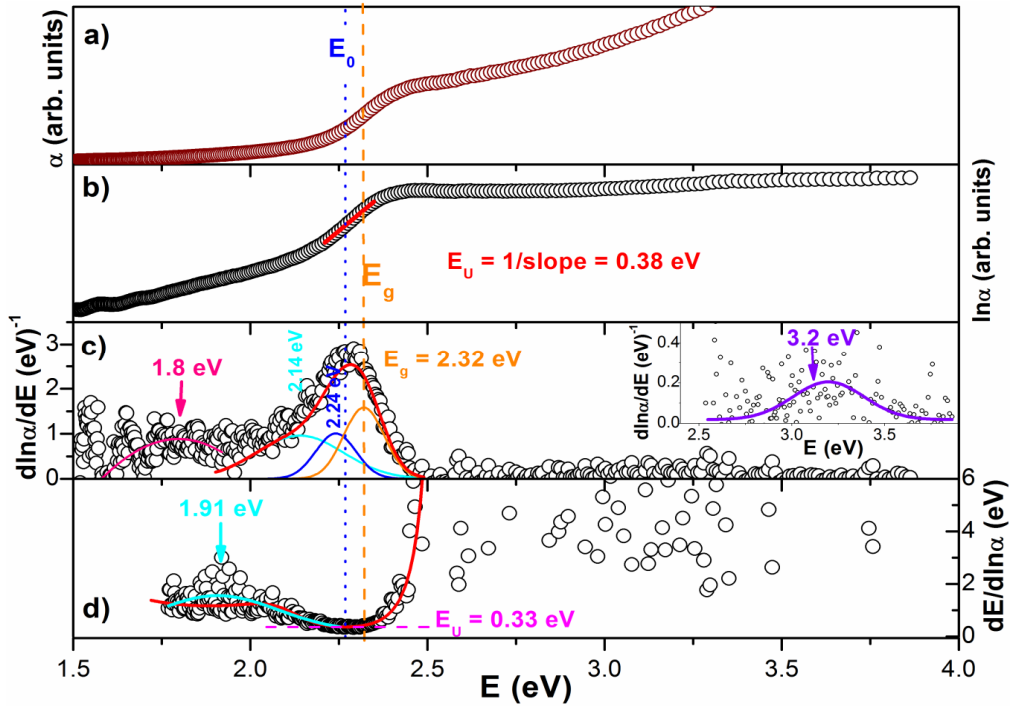

**Fig. 5** The energy-dependent curves of a) absorption coefficient ($\alpha$) of oxygen-deficient BiFeO$_{2.85}$ thick film. b) its logarithmic term, $\ln\alpha$, c) its PDF function, $\left(\frac{d\ln\alpha(E)}{dE}\right)$, and d) the corresponding CDF function, $\left(\frac{dE}{d\ln\alpha(E)}\right)$.

In addition, three distinct peaks at 1.8 eV, 2.14 eV, and 2.24 eV were observed, which correspond to three of the expected four *d-d* transitions of Fe$^{3+}$ ions in the BiFeO$_{2.85}$ film, in accordance with the literature [14,26,34,35]. These values are also in excellent agreement with the energies of the *d-d* transitions of $Fe^{3+}$ ions in bulk and nanostructured $BiFeO_3$ samples (see Figs. 2 and 4). Additionally, the CDF data (Fig. 5d) also clearly shows a peak feature at 1.91 eV, which is likely caused by an indirect transition in the $BiFeO_{2.85}$ film. The slope drawn at the inflection point of the Urbach focus yielded an $E_U$ value of 0.33 eV (Table 1) for the $BiFeO_{2.85}$ thick film, which is comparable to the value ($E_U$ = 0.38 eV) determined using the Urbach relation in equation (2).

### *3.4 The electronic band structure based on the obtained energies of electronic transition in the studied $BiFeO_3$-based materials*

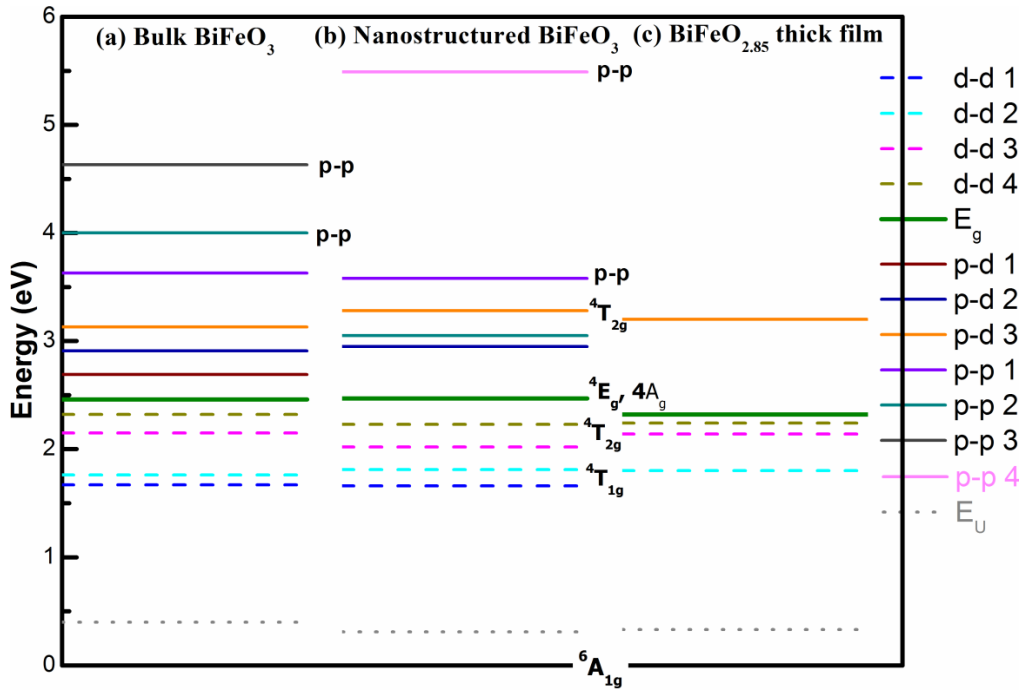

**Fig. 6** The obtained electronic transition energies for the samples a) bulk $BiFeO_3$, b) nanostructured $BiFeO_3$, and c) $BiFeO_{2.85}$ thick film using the analysis of the PDF and CDF functions derived from their experimental absorption spectra.

Figure 6 displays the electronic band structures for the three $BiFeO_3$-based materials that were examined. This was achieved by utilizing the transition energies of the allowed *d-d*, *p-d*, and *p-p* charge transfer transitions of bulk $BiFeO_3$, nanostructured $BiFeO_3$, and $BiFeO_{2.85}$ thick film obtained through the analysis of their PDF and CDF functions. Specifically, three transitions associated with the *p-d* charge transfer transitions caused by the electron-hole recombination process in octahedral $Fe^{3+}O_6$ centers at 2.69 eV, 2.91 eV, and 3.16 eV were detected for bulk $BiFeO_3$ [26,53,54]. Similarly, three *p-d* charge transfer transitions were observed at 2.95 eV, 3.05 eV, and 3.28 eV for nanostructured $BiFeO_3$ (Fig. 6b) and only one *p-d* charge transfer transition was witnessed at 3.20 eV for the $BiFeO_{2.85}$ film (Fig. 6c). Conversely, three transitions related to the *p-p* charge transfer transitions due to the O 2p-Bi 4p hybridization at 3.63 eV, 4.0 eV, and 4.63 eV were observed for bulk $BiFeO_3$ [53]. While there were two *p-p* charge transfer transitions at 3.58 eV and 5.49 eV for nanostructured $BiFeO_3$, there was no indication of the *p-p* transitions for the $BiFeO_{2.85}$ film. The energy values of the observed *p-d* and p-p charge transfer transitions in the $BiFeO_3$-based materials correlate well with the experimental and theoretical findings in the literature [12-14,26,35,53,54].

Importantly, four transitions associated with two doubly degenerate *d-d* transitions of $Fe^{3+}$ ions in $FeO_6$ octahedra were distinctly detected at 1.67 eV, 1.76 eV, 2.15 eV, and 2.32 eV (indicated by dashed lines in Fig. 6a) due to the low-symmetry crystal field, spin-orbit and electron-lattice interaction in $BiFeO_3$ [26-28,34,53]. Similarly, the nanostructured $BiFeO_3$ sample showed four *d-d* transitions at 1.66 eV, 1.81 eV, 2.02 eV, and 2.23 eV (Fig. 6b), while the $BiFeO_{2.85}$ film only showed three *d-d* transitions at 1.80 eV, 2.14 eV, and 2.24 eV (Fig. 6c). Notably, the direct band gap transition for bulk and nanostructured $BiFeO_3$ samples was observed at 2.46 eV and 2.47 eV (see Table 1), illustrated as olive solid lines in Fig. 6a and Fig. 6b, respectively. The direct band gap energy of the $BiFeO_{2.85}$ thick film was approximately 2.32 eV (see Fig. 6c and Table 1), which is predictably red-shifted compared to bulk $BiFeO_3$ [35]. Interestingly, the weak indirect transition was detected near 1.90 eV for all three investigated $BiFeO_3$-based samples (which are not shown here), consistent with the reported values (1.70-1.93 eV) of these materials obtained using the Tauc plot analysis [34,35]. In summary, the current investigation on the electronic transitions and electronic band structure of bulk $BiFeO_3$, nanostructured $BiFeO_3$, and $BiFeO_{2.85}$ thick film was conducted using a unique method known as derivative absorption spectroscopy analysis. The evaluated energies of the various electronic transitions (see Figs. 1–6) are in excellent agreement with the results using conventional methods found in the literature [12-14,26,35,53,54]. Because of this, the implemented derivative absorption spectroscopy analysis method can be used to probe the complex electronic structure of strongly correlated electronic compounds like the magnetoelectric multiferroic $BiFeO_3$.

## 4. Conclusion

This study employed a unique technique that combined derivative spectroscopic analysis with a deconvolution approach to comprehensively analyze the absorption spectra of bulk, nanostructured, and film samples of the magnetoelectric multiferroic $BiFeO_3$. Within the energy ranges of 1.5–2.5 eV, 2.5–3.2 eV, and 3.2–5.0 eV, three distinct types of electronic transitions were observed in all three materials studied: dipole-allowed *d-d* transitions, *p-d* charge transfer transitions, and *p-p* charge transfer transitions, respectively. Additionally, all samples exhibited a weak indirect transition alongside a strong direct transition linked to the optical band-edge of $BiFeO_3$. In the $BiFeO_{2.85}$ film, the direct band gap energy ($E_g$ = 2.32 eV) was expectedly red-shifted compared to bulk $BiFeO_3$ ($E_g$ = 2.46 eV). One intriguing aspect of this analysis was the direct evaluation of the energy of the defect-induced Urbach tail bands. This was achieved by estimating the slope at the inflection point of the Urbach focus within the calculated cumulative density function (CDF, $\frac{dE}{d \ln \alpha(E)}$) from the absorption spectra of the materials. Using this novel method, the Urbach energy for bulk, nanostructured, and film samples of $BiFeO_3$ was directly calculated, yielding values of approximately 0.40 eV, 0.31 eV, and 0.33 eV, respectively. These values are in good agreement with the values (0.35–0.48 eV) obtained from the Urbach empirical formula for these materials. Finally, the obtained energies of electronic transitions were plotted for all three inspected materials to schematize a comprehensive electronic band structure for them.